\documentclass[aip,twocolumn,apl,reprint,amsmath,amssymb,superscriptaddress]{revtex4-1}

\usepackage{graphicx}% Include figure files

\begin{document}

\title{Random telegraph noise in metallic single-walled carbon nanotubes}

\author{Hyun-Jong Chung}
\author{Tae Woo Uhm}
\author{Sung Won Kim}
\author{Young Gyu You}
\author{Sang Wook Lee}
\affiliation{Division of Quantum Phases and Devices, School of Physics, Konkuk University, Seoul 143-701, Korea}
\author{Sung Ho Jhang}
\altaffiliation{Author to whom correspondence should be addressed. electronic mail: shjhang@konkuk.ac.kr}
\affiliation{Division of Quantum Phases and Devices, School of Physics, Konkuk University, Seoul 143-701, Korea}

\author{Eleanor E. B. Campbell}
\affiliation{Division of Quantum Phases and Devices, School of Physics, Konkuk University, Seoul 143-701, Korea}
\affiliation{EaStCHEM, School of Chemistry, Edinburgh University, West Mains Road, Edinburgh EH9 3JJ, United Kingdom}

\author{Yung Woo Park}
\affiliation{Department of Physics and Astronomy, Seoul National University, Seoul 151-747, Korea}

\date{\today}

\begin{abstract}
We have investigated random telegraph noise (RTN) observed in individual metallic
carbon nanotubes (CNTs). Mean lifetimes in high-
and low-current states, $\tau_{\text{high}}$ and
$\tau_{\text{low}}$, have been studied as a function of
bias-voltage and gate-voltage as well as temperature. By analyzing
the statistics and features of the RTN, we suggest that this noise
is due to the random transition of defects between two metastable
states, activated by inelastic scattering with conduction
electrons. Our results indicate an important
role of defect motions in the $1/f$ noise in CNTs.
\end{abstract}

\pacs{73.63.Fg, 72.70.+m, 73.23.-b}

\maketitle

%\section{Introduction}
The switching of resistance between two discrete values, referred
to as random telegraph noise \cite{Review}, has been
observed in a variety of mesoscopic systems such as submicron
metal-oxide-semiconductor field-effect-transistors (MOSFETs)
\cite{Ralls,Grasser}, metallic nanobridges \cite{Ralls2}, and small tunnel
junctions \cite{Farmer}. Although microscopic details differ from
one system to another, the observed switching of resistance is an
apparent signature of an underlying two-level fluctuator (TLF),
which consists of two energy wells separated by a barrier. The
presence of a large number of TLF's, with a wide distribution of
fluctuation rates, is generally believed to be responsible for the
1/\textit{f} noise, frequently observed in various materials and
systems. 1/\textit{f} noise
has also been widely studied in carbon nanotubes
\cite{Zettl,Hakonen1,Hakonen2,Dekker,Roumiantsev,Roche,Ouacha,Lin},
and the random telegraph signal has been reported for CNT-FETs\cite{Liu1,Liu2,Liu3}
and CNT film-silicon Schottky junctions\cite{An}.
These RTNs were attributed to charge traps in dielectric materials or in the interface.

In this letter, we report extensive
observations of RTN in individual metallic CNTs. The noise behavior is distinguished from the RTN observed in semiconducting CNT-FETs\cite{Liu1,Liu2}.
By analyzing the statistics and features of the current switching, we attribute the
RTN to the defect motions between two metastable states. The
activation energy for this transition is evaluated from the
bias-voltage dependence of the RTN.

\begin{figure}[b]
\includegraphics[width=7.7cm]{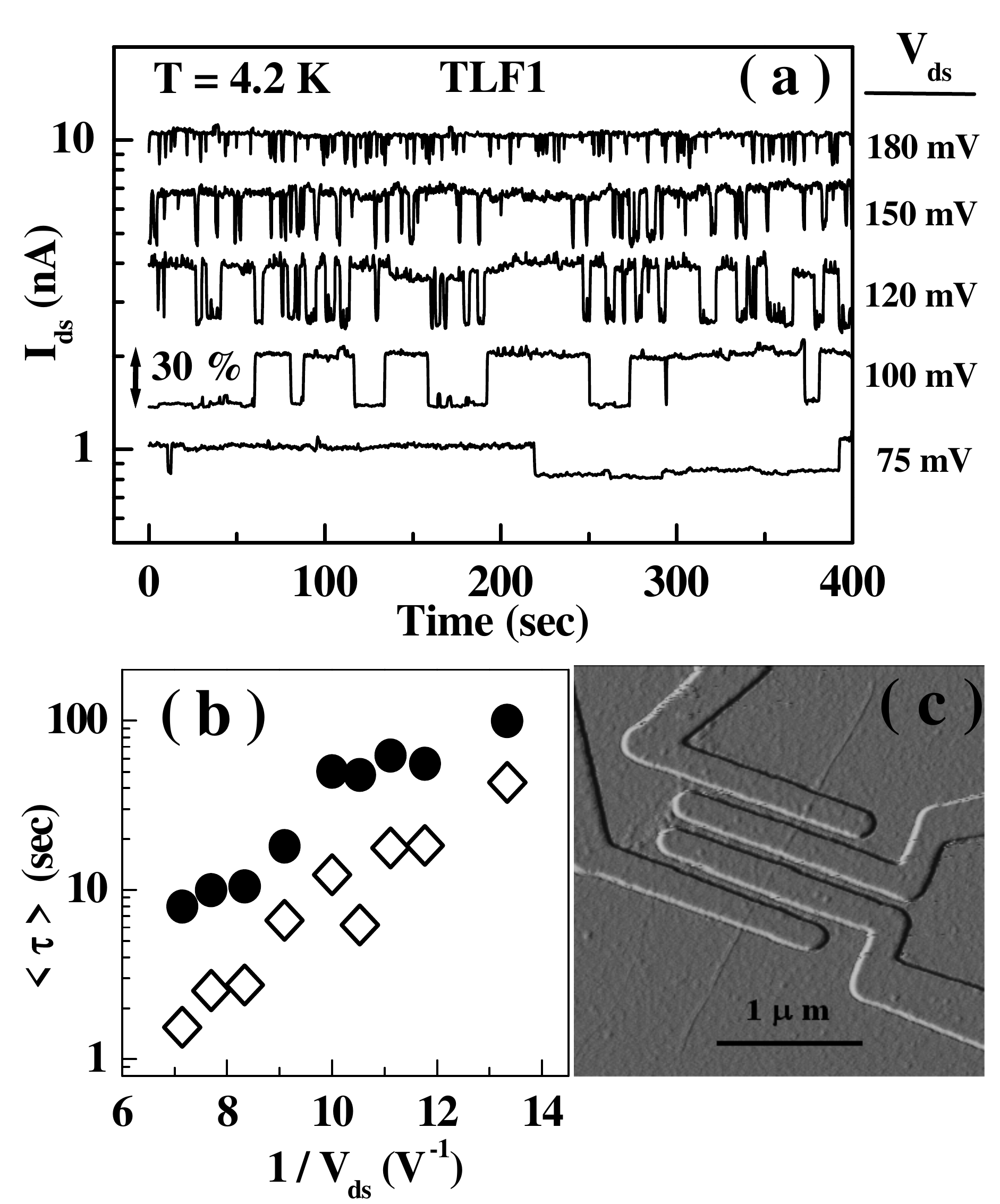}
\caption{\label{fig:epsart} Random telegraph noise observed in a
metallic SWNT at \textit{T} = 4.2 K. (a) Time-traces of currents
for five different $V_{\text{ds}}$. Fluctuation rate becomes
faster with increasing $V_{\text{ds}}$. The magnitude of current
fluctuation reaches 30$\%$ of total current. (b) Exponential
dependence of mean lifetimes on inverse $V_{\text{ds}}$:
$\tau$$_{\text{high}}$ [$\bullet$] and $\tau$$_{\text{low}}$
[$\diamond$]. (c) Typical tapping-mode atomic force microscope
image of SWNT with Ti/Au electrodes on it.}
\end{figure}

%\section{Experimental procedure}
Experiments have been carried out on individual metallic single-walled carbon nanotubes (SWNTs)
dispersed on Si/SiO$_{2}$ substrates. The heavily doped Si was
used as a back-gate and the thickness of the oxide layer was 300~nm.
For electrical contacts, Ti/Au (5~nm/15~nm) electrodes were
deposited on SWNTs using conventional e-beam lithography (Figure 1
(c)). Low temperature measurements were performed both in a
Janis variable temperature cryogenic system and in a simple liquid
He bath. The samples were biased at a constant voltage and the
current fluctuations were monitored with either a preamplifier
(Ithaco~1211) or a semiconductor characterization system
(Keithley~4200). In general, the RTN can be characterized by three
parameters, namely the RTN amplitude
($\Delta$\textit{I}$_{\text{ds}}$) and the mean lifetimes of the
 high-current state ($\tau$$_{\text{high}}$) and the low-current
state ($\tau$$_{\text{low}}$). To obtain reasonable
statistical values of these parameters, 5000 to 20000 current
points were registered for a fixed drain-source bias-voltage
($V_{\text{ds}}$). The observation window of
$\tau$$_{\text{high}}$, $\tau$$_{\text{low}}$ lies between
0.1 s and 1000 s.

\begin{figure}[t]
\includegraphics[width=7.5cm]{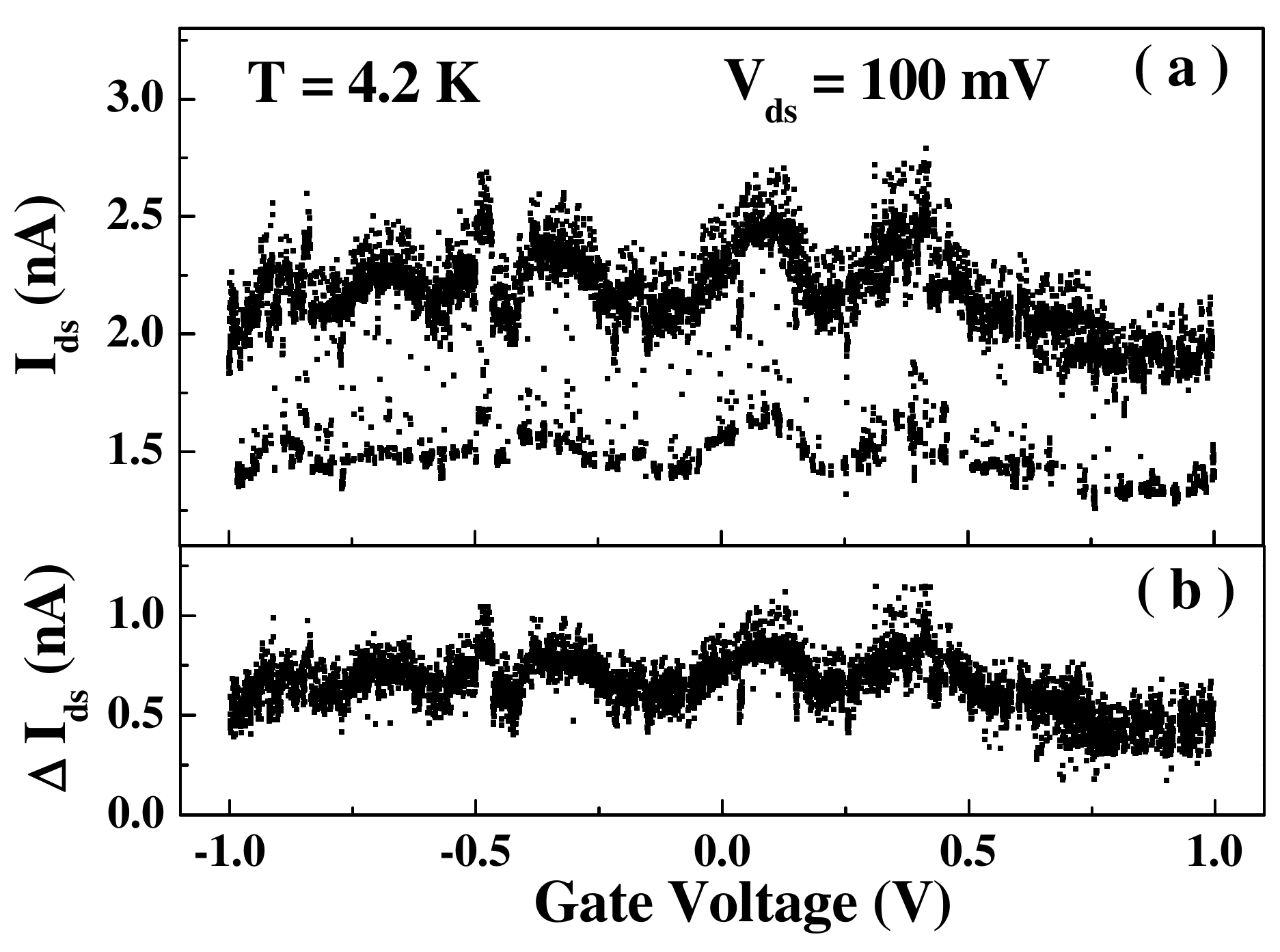}
\caption{\label{fig:epsart} (a) Drain current as a function of
gate-voltage at \textit{T} = 4.2 K while keeping $V_{\text{ds}}$ =
100 mV. Two discrete current levels are clearly observed in
Coulomb oscillations. (b) RTN amplitude ($\Delta$$I_{\text{ds}}$)
calculated from Fig.\ 2(a). $\Delta$$I_{\text{ds}}$ shows
identical peaked features with $I_{\text{ds}}$ presenting maximum
noise amplitude at Coulomb conductance peaks.}
\end{figure}

%\section{Results and Discussion}
In Figs.\ 1 and 2, we present typical results of two-probe
measurements from a representative sample. Figure 1(a) shows time-traces of the
drain-source currents ($I_{\text{ds}}$) for five different
$V_{\text{ds}}$ at \textit{T} = 4.2 K. Current switching between
two discrete values is clearly observable in a particular range of
$V_{\text{ds}}$, 75 $\leq$ $V_{\text{ds}}$ $\leq$ 180 mV, and the
magnitude of the current fluctuation reaches 30$\%$ of the total
current. The fluctuation rate becomes faster with increasing
$V_{\text{ds}}$ and the switching becomes faster than the
experimental bandwidth for $V_{\text{ds}}$ $\geq$ 180 mV. Values
of $\tau$$_{\text{high}}$ ($\bullet$) and $\tau$$_{\text{low}}$
($\diamond$), obtained from nine different $V_{\text{ds}}$, are
shown in Fig.\ 1(b). It demonstrates that the
$\tau$$_{\text{high}}$ and $\tau$$_{\text{low}}$ increase
exponentially with respect to the inverse $V_{\text{ds}}$. To
investigate the effect of gate-voltage ($V_{g}$) on the RTN, we
checked the current fluctuation as we swept $V_{g}$ by 20 $\mu$V
per second in the range of 1V while keeping $V_{\text{ds}}$ = 100
mV. Figure 2(a) shows the measured currents as a function of
gate-voltage. The two current levels are clearly distinguishable
in the Coulomb oscillations. In Fig.\ 2(b), the noise amplitude
$\Delta$$I_{\text{ds}}$ was estimated by taking the difference of
the two discrete current-curves in Fig.\ 2(a). The peak positions
of $\Delta$$I_{\text{ds}}$ match with those of $I_{\text{ds}}$.
The result is consistent with the reported 1/\textit{f} noise
characteristics of a SWNT single-electron-transistor (SET)
\cite{Dekker}, where peaks of both the current and the current
noise coincide.

\begin{figure}[b]
\includegraphics[width=8cm]{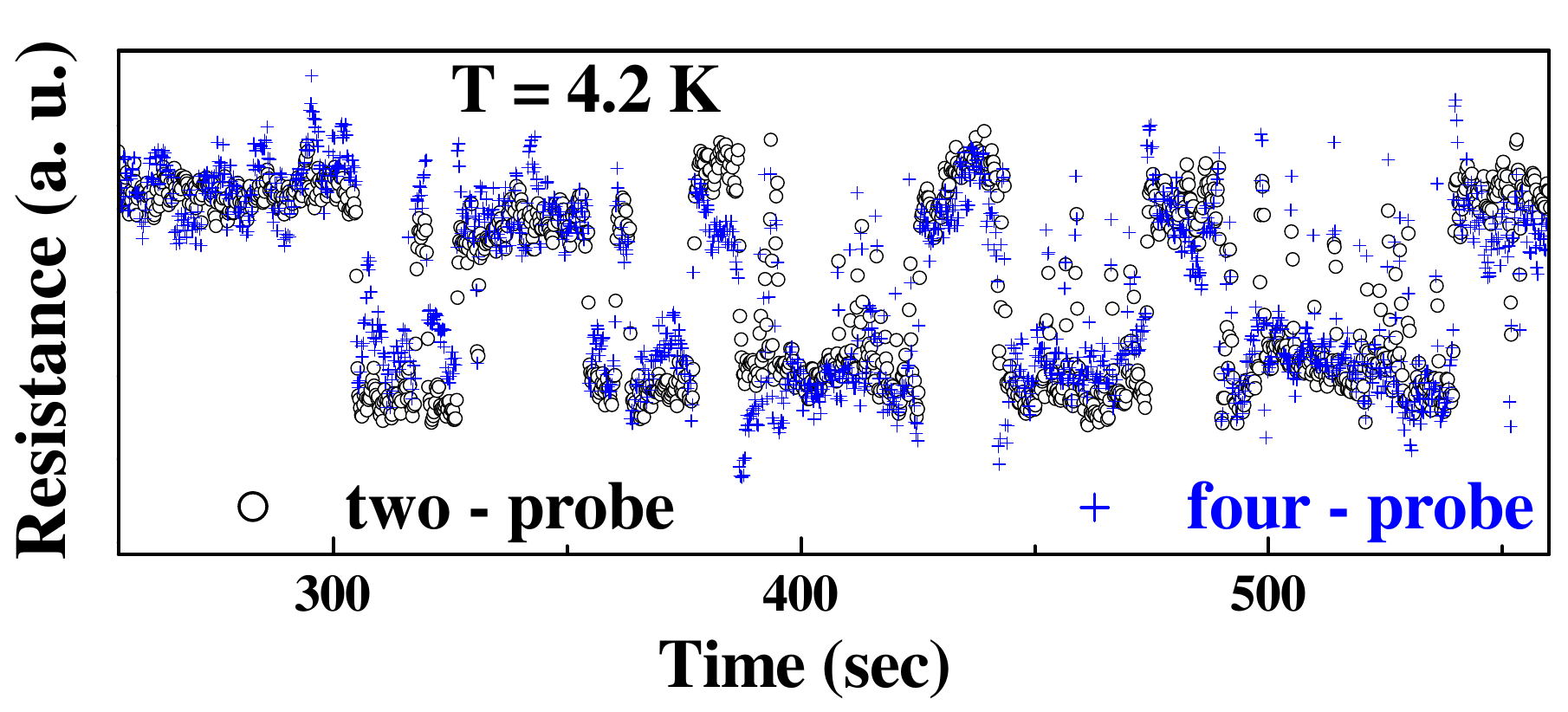}
\caption{\label{fig:epsart} Resistance traces from four- and
two-probe measurements in arbitrary units. The two traces match
each other well, proving that the RTN is not due to the contact
barriers. For eye convenience, both resistance traces, which show
different values, are shifted into the same place.}
\end{figure}

What is the origin of the RTN we have observed in metallic SWNTs?
At first, we checked the effect of tunnel conductance fluctuations
across the contact barrier, which can possibly cause the RTN,
assuming two kinds of contact configurations with different tunnel
barriers. To test this, a four-probe measurement was introduced
with the electrode configuration shown in Fig.\ 1(c). We
simultaneously measured the RTN in both four- and two-probe
configurations, i.e., measuring the $I_{\text{ds}}$ and
$V_{\text{ds}}$ between the outer two electrodes (two-probe) and
at the same time measuring the voltage drop between the inner two
electrodes (four-probe) of the same sample. Figure 3 shows that
both the two- and four-probe resistance switch at the same time,
which rules out the role of contact barriers in the RTN.

As a second candidate, we can possibly think about the effect of
charge traps in dielectric materials which could explain the RTN
in MOSFETs \cite{Ralls,Grasser} and CNT-FETs \cite{Liu1,Liu2,Liu3}, since our experiments have been performed
for similar FET structures. However, our results of the RTN in metallic SWNTs are different
from those observed in the semiconductors. The same experiment as
in Fig.\ 2(a), performed on a submicron MOSFET in the
Coulomb-blockade regime at $T=4.2$~K, showed two discrete
$I_{\text{ds}}$-$V_{g}$ curves that exhibited the same Coulomb
oscillations but were shifted relative to one another along the
horizontal axis ($V_{g}$) \cite{Peters}. This is because the
trapped charge affects the potential of the dot.
In our RTN data, however, no horizontal shift is observed as shown in
Fig.\ 2(a). Furthermore, the noise peaks, shown in Fig.\ 2(b),
occur at the zero-gain points,
($\partial$$I_{\text{ds}}$/$\partial$$Q_{g}$ = 0), where the
current is a maximum. If the noise is caused by charge
fluctuations the noise peak should occur at the point of maximum
gain, that is, at $V_{g}$ where $I_{\text{ds}}$ is most sensitive
to slight fluctuations in gate-voltage \cite{Delsing,Hakonen1}.
Secondly, $\tau$$_{\text{high}}$ and $\tau$$_{\text{low}}$ showed
no gate-voltage dependence in the range of $V_{g}$ = $-16$ to 16 V
(data not shown). However, a negatively charged trap in an oxide
layer is more likely to emit its charge at $V_{g}$ $\ll$ 0 and to
maintain it at $V_{g}$ $\gg$ 0.
Note that the RTNs in CNT-FETs are observed only in a limited range of the gate-voltage \cite{Liu1,Liu2}.
Therefore, the lack of a
gate-voltage dependence of $\tau$$_{\text{high}}$ and
$\tau$$_{\text{low}}$ indicates that the RTN is not due to charge
fluctuations. Finally, to rule out the effect of traps in the
substrate, we prepared a SWNT suspended over the SiO$_{2}$
substrate \cite{nicesw}. The RTN still appeared in the suspended
SWNT at \textit{T} = 1.8 K (data not shown). Therefore, we
conclude that it is necessary to consider another source of the
RTN.

Regarding the RTN to be due to intrinsic fluctuations, we now turn
our attention to the defects in metallic SWNTs. Considering the
high current density ($\sim$10$^{6}$ A/cm$^{2}$ at 10 nA) flowing
through the surface of a 1 nm-sized carbon tube, a defect in a
SWNT could transfer between two metastable positions, activated by
inelastic scattering with conduction electrons. The reversible
motion of a defect, inducing different electrical properties at
each metastable position, could produce the observed telegraphic
current fluctuations. We note that RTN in metallic nanobridges has
been successfully explained in terms of similar defect motion
\cite{Ralls2,Ralls3,Ralls4,Holweg,Muller}. This noise mechanism is
consistent with our typical observation of electromigration of
defects, which appears as an irreversible current change in time
as $V_{\text{ds}}$ is further increased.

For the analysis of our results, we adopt the model
\cite{Ralls3,Holweg} used to describe RTN in metallic nanobridges.
The temperature of a defect is usually identical to the lattice temperature.
However, in CNTs and metallic nanobridges, the defect temperature $T_{d}$ is expected to be much higher than the lattice temperature
because of the inelastic scattering with conduction electrons as well as the poor energy relaxation to the lattice.
With $T_{d}$ depending on the bias voltage,
the model could explain the exponential dependence of the mean
lifetimes as a function of $V_{\text{ds}}^{-1}$, observed at large
bias-voltages \cite{Ralls3,Holweg,Muller}. Note that we found a
similar dependence in Fig.\ 1(b).
Following the approach of Ref.\ \cite{Ralls3}, where they
calculated $T_{d}$ in equilibrium with ballistic electrons
(treating the defect as a harmonic oscillator), Holweg \textit{et
al}.\ derived the relation $k_{B}$$T_{d}$ =
$\alpha$\textit{e}$|$$V_{\text{ds}}$$|$ with $\alpha$ = 5/16 for
high bias-voltage and low lattice temperature \cite{Holweg}. With
a modification term due to the electromigration force, the
thermally activated behavior of the mean lifetime $\tau$ either in
the high- or low-current state was expressed by
\begin{eqnarray*}
\tau = \tau_{0} \exp
 \left(
  \frac{E_{B} - \zeta V_{\text{ds}}}{\alpha e|V_{\text{ds}}|}
 \right)
\end{eqnarray*}
with $\tau_{0}$ the attempt time, $E_{B}$ the activation energy,
and $\zeta$ the electromigration parameter. From the slope of
Fig.\ 1(b) the activation energy $E_{B}$ of TLF1 in Fig.\ 1(a) is
estimated (with $\alpha$ = 5/16) to be 140~meV for
$\tau$$_{\text{high}}$ and 160~meV for $\tau$$_{\text{low}}$.
However, here we point out that for some TLF the rate becomes
independent of $V_{\text{ds}}$ at low bias-voltage as shown in
Fig.\ 4. Also, the temperature dependence of the RTN, displayed in
the insert of Fig.\ 4, shows that the fluctuation rate is nearly
independent of temperature at \textit{T} $\leq$ 20 K, indicating
that tunnelling between the two metastable states, rather than
thermal activation, is dominant in this temperature range. Based
on these observations, we assume that $T_{d}$ =
$\alpha$$e|V_{\text{ds}}|$/$k_{B}$ is equal to 20 K at
$V_{\text{ds}}^{-1}$ = 19.5~V$^{-1}$ where the fluctuation rate
becomes saturated at low bias-voltage. Thus we obtain $\alpha$
$\sim$ 0.034 for the TLF2 in Fig.\ 4. This value is an order of
magnitude smaller than $\alpha$ = 5/16 suggested by the theory for
metallic nanobridges. The voltage drop at the contact between the
SWNT and the electrodes could be responsible for the reduced value
of $\alpha$, together with the effect of energy relaxation to the
lattice, which is not accounted for in the above theory. Also, the
small $\alpha$ parameter reflects our diffusive SWNT device, allowing an electron to scatter off defects
several times while traversing the tube.

\begin{figure}[t]
\includegraphics[width=8cm]{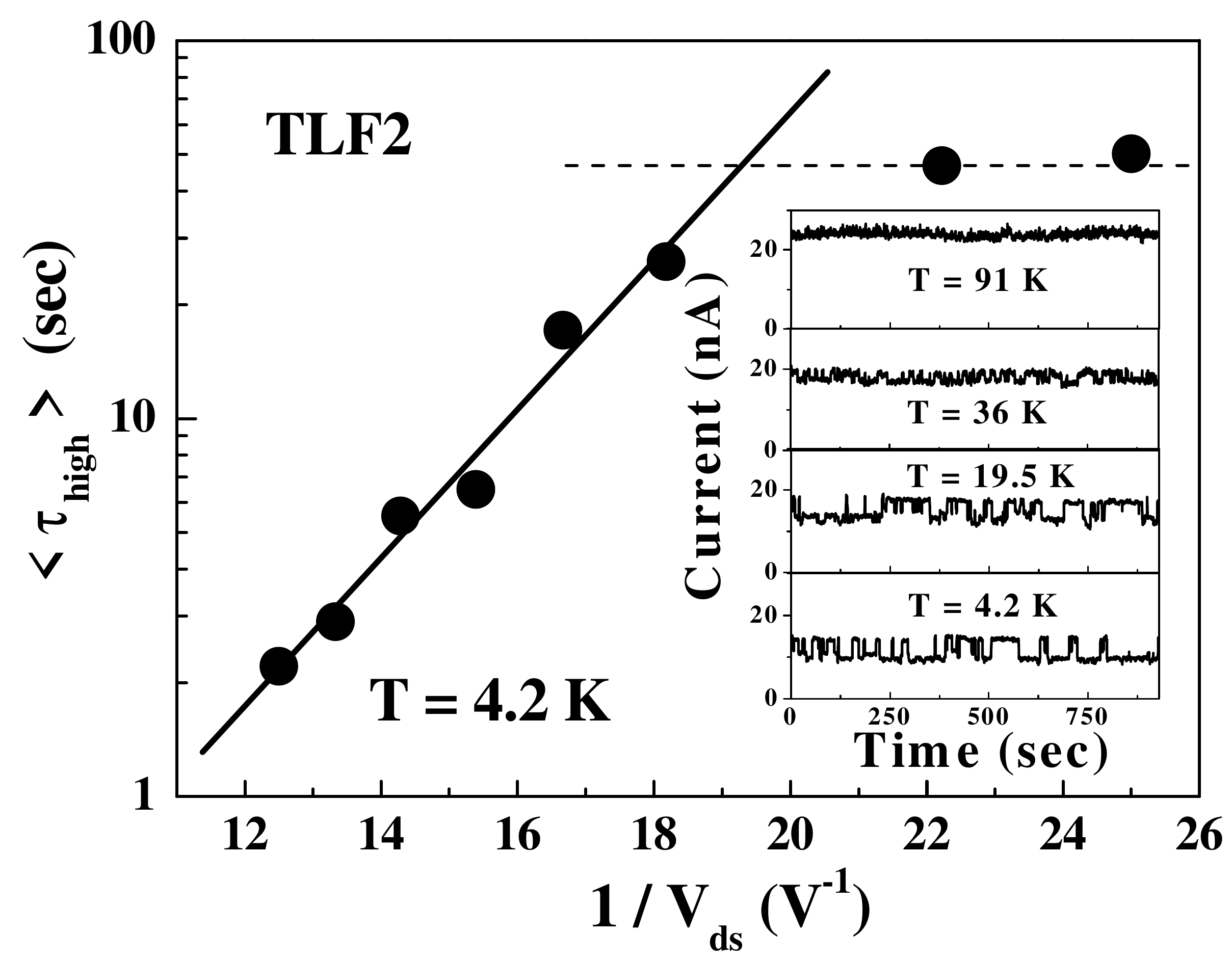}
\caption{\label{fig:epsart} $\tau$$_{\text{high}}$ obtained from
eight different $V_{\text{ds}}$ for a SWNT at \textit{T} = 4.2 K.
While again the exponential dependence on inverse $V_{\text{ds}}$
is observed, the fluctuation rate becomes independent of
$V_{\text{ds}}$ at low bias-voltage. Insert displays the
temperature dependence of the RTN measured at $V_{\text{ds}}$ = 50
mV. Fluctuation rate is nearly independent of temperature for
\textit{T} $\leq$ 20 K, indicating that tunnelling between the two
metastable states is dominant in this temperature range.}
\end{figure}

In Table I, the magnitude of the current fluctuation and the
estimated activation energy (with experimentally determined
$\alpha$) are summarized for three different TLF's.
The measured $\Delta$$I_{\text{ds}}$/$I_{\text{ds}}$ $\sim$
0.1--0.3 is two to four orders of magnitude larger than that of
the metallic nanobridges (diameter $\sim$ 10~nm)
\cite{Ralls2,Ralls3,Ralls4,Holweg}. This can be attributed to the
much narrower current path in the SWNTs. In fact, for atomic-scale
metal-constriction, the magnitude of the fluctuation can be as
large as the total conductance \cite{Muller}. The small activation
energy of 15--24~meV is a reflection of our measurement
temperature (\textit{T} = 4.2~K),
and it is comparable to those of the
metallic nanobridges measured at the same temperature.
In most systems where RTN was found, the measured activation energy was a
strong function of the temperature \cite{Ralls,Ralls4}. This is
because only the TLF for which the activation energy corresponds
to the measurement temperature can be observable as RTN in the
experimental bandwidth. At higher temperatures, the RTN was often
observed with several TLF's acting at the same time. In that case,
the frequency dependence of the noise became close to the 1/$f$
spectrum. If we regard the 1/$f$ noise as a superposition of such
TLF's, our results indicate an important role of defect motions in
the 1/$f$ noise observed in the CNTs
\cite{Zettl,Hakonen1,Hakonen2,Dekker,Roumiantsev,Roche,Ouacha}. In
many CNT devices, prepared on dielectric substrates, charge traps
in the vicinity of CNTs are expected to play a role in the 1/$f$
noise. However, the deviation from the typical gain dependence of
the 1/$f$ noise, reported for CNT-SETs \cite{Hakonen1,Dekker},
cannot be explained by the charge fluctuations alone and instead
can be understood by invoking a noise mechanism due to the defect
motions. Also, in the frequency domain, the current power spectral
density of the RTN is a Lorentzian given by \cite{Machlup}
\begin{eqnarray*}
\frac {S_{I}(f)}{I_{\text{ds}}^{2}} =
  \frac{4 (\Delta I_{\text{ds}}/I_{\text{ds}})^{2}}{(\tau_{\text{high}}
  + \tau_{\text{low}})[( 1 / \tau_{\text{high}} +1 / \tau_{\text{low}} )^{2}+ (2 \pi
  f)^{2}]}.
\end{eqnarray*}
With the 1/$f^{2}$ tail of the Lorentzian, we note that the
1/$f^{2}$ dependence of the noise (instead of 1/$f$) observed in
free-standing CNTs \cite{Hakonen2,Roche} can be interpreted as due
to the presence of RTN, generated by defect motions.

Large RTN amplitude observed for metallic SWNTs suggests the
possibility to use RTN measurements as a sensitive probe for
characterizing the defects in nanotubes. Also, it is remarkable
that, increasing the number of defects on the tubes by Cs ion
irradiation, we could sometimes observe more current levels
appearing in the time traces, which resulted in overall higher
resistance fluctuations. With further investigations, RTN approach
could be developed into a comparative diagnostic to grade
differently prepared nanotubes.

\begin{table}
\caption{\label{tab:table1} The magnitude of current fluctuation
and the activation energy summarized for three different TLF's
observed in metallic SWNTs \cite{alpha}.
 }
\begin{ruledtabular}
\begin{tabular}{ccccc}
Fluctuator & $\Delta$$I_{\text{ds}}$/$I_{\text{ds}}$ & Current State &
$E_{B}$
(meV) & $\alpha$\\
\hline
TLF1 & 0.3 & high &20.7 & \\
& & low & 24 \\
TLF2 & 0.33 & high & 15.3 & 0.034\\
 & & low & 15.6 & \\
TLF3 & 0.1 & low & 15.4 & 0.058\\
\end{tabular}
\end{ruledtabular}
\end{table}

In summary, we have investigated random telegraph noise observed
in individual metallic SWNTs. Reversible motion of a defect,
activated by inelastic scattering with conduction electrons, is
suggested to be responsible for the observed RTN. Regarding the
1/$f$ noise as a superposition of two-level current switchings,
our results imply an important role of defect motions as a source
of the 1/$f$ noise for the CNTs.

%\begin{acknowledgments}
This paper was supported by Konkuk University in 2013.
%\end{acknowledgments}


\begin{thebibliography}{}

\bibitem{Review} For an extensive review, see Sh. Kogan, \textit{Electronic noise
and fluctuations in solids} (Cambridge Univ. Press, Cambridge,
1996), Chap. 8; M. J. Kirton and M. J. Uren, Adv. Phys. {\bf 38},
367 (1989).
\bibitem{Ralls} K. S. Ralls, W. J. Skocpol, L. D. Jackel, R. E. Howard, L. A. Fetter, R. W. Epworth and D. M. Tennant, Phys. Rev. Lett. {\bf 52}, 228 (1984).
\bibitem{Grasser} T. Grasser, Microelectronics Reliability {\bf 52}, 39 (2012).
\bibitem{Ralls2} K. S. Ralls and R. A. Buhrman, Phys. Rev. Lett. {\bf 60}, 2434 (1988).
\bibitem{Farmer} K. R. Farmer, C. T. Rogers, and R. A. Buhrman, Phys. Rev. Lett. {\bf 58}, 2255 (1987).
\bibitem{Zettl} P. G. Collins, M. S. Fuhrer, and A. Zettl, Appl. Phys. Lett. {\bf 76}, 894 (2000).
\bibitem{Hakonen1} M. Ahlskog, R. Tarkiainen, L. Roschier, and P. Hakonen, Appl. Phys. Lett. {\bf 77}, 4037 (2000).
\bibitem{Hakonen2} L. Roschier, R. Tarkiainen, M. Ahlskog, M. Paalanen, and P. Hakonen, Appl. Phys. Lett. {\bf 78}, 3295 (2001).
\bibitem{Dekker} H. W. Ch. Postma, T. F. Teepen, Z. Yao, and. C. Dekker, in \textit{Electronic correlations: from
meso- to nano-physics}, edited by Th. Martin and G. Montambaux
(EDP Sciences, France, 2001).
\bibitem{Roumiantsev} R. Vajtai, B. Q. Wei, Z. J. Zhang, Y. Jung, G. Ramanath, and P. M. Ajayan, Smart Mater. Struct. {\bf 11}, 691 (2002).
\bibitem{Roche} P.-E. Roche, M. Kociak, S. Gueron, A. Kasumov, B. Reulet, and H. Bouchiat, Eur. Phys. J. B {\bf 28}, 217 (2002).
\bibitem{Ouacha} H. Ouacha, M. Willander, H. Y. Yu, Y. W. Park, M. S. Kabir, S. H. M. Persson, L. B. Kish, and A. Ouacha, Appl. Phys. Lett. {\bf 80}, 1055 (2002).
\bibitem{Lin} Y.-M. Lin, J. Appenzeller, Z. Chen, and P. Avouris, Physica E {\bf 37}, 72 (2007).
\bibitem{Liu1} F. Liu, M. Bao, H. J. Kim, K. L. Wang, C. Li, X. Liu, and C. Zhou, Appl. Phys. Lett. {\bf 86}, 163102 (2005).
\bibitem{Liu2}  F. Liu, K. L. Wang, D. Zhang, and C. Zhou, Appl. Phys. Lett. {\bf 89}, 243101 (2006).
\bibitem{Liu3} F. Liu and K. L. Wang, Nano Lett. {\bf 8}, 147 (2008).
\bibitem{An} Y. An, H. Rao, G. Bosman, and A. Ural, Appl. Phys. Lett. {\bf 100}, 213102 (2012).
\bibitem{Peters} M. G. Peters, J. I. Dijkhuis, and L. W. Molenkamp, J. Appl. Phys. {\bf 86}, 1523 (1999).
\bibitem{Delsing} B. Starmark, T. Henning, T. Claeson, and P. Delsing, J. Appl. Phys. {\bf 86}, 2132 (1999).
\bibitem{nicesw} S. W. Lee, D. S. Lee, H. Y. Yu, E. E. B. Campbell, and Y. W. Park, Appl. Phys. A {\bf 78}, 283 (2004).
\bibitem{Ralls3} K. S. Ralls, D. C. Ralph, and R. A. Buhrman, Phys. Rev. B {\bf 40}, 11561
(1989).
\bibitem{Ralls4} K. S. Ralls and R. A. Buhrman, Phys. Rev.
B {\bf 44}, 5800 (1991).
\bibitem{Holweg} P. A. M. Holweg, J. Caro, A. H. Verbruggen, and S.
Radelaar, Phys. Rev. B {\bf 45}, 9311 (1992).
\bibitem{Muller} C. J. Muller, J. M. van Ruitenbeek, and L. J. de Jongh,
Phys. Rev. Lett. {\bf 69}, 140 (1992).
\bibitem{alpha} Since we could not measure the $\alpha$ at the low
$V_{\text{ds}}$ range for TLF1, we simply took the average value
of those for TLF2 and TLF3.
\bibitem{Machlup} S. Machlup, J. Appl. Phys. {\bf 25}, 341 (1954).
\end{thebibliography}
\end{document}